\documentclass[aps,pre,groupedaddress,twocolumn,10pt]{revtex4-1}
\usepackage{graphicx}
\usepackage[english]{babel}
\usepackage[T1]{fontenc}

\usepackage{amsmath}
\usepackage{amssymb}
\usepackage{amsthm}
\usepackage{bbm}
\usepackage{hyperref}
\usepackage[T1]{fontenc}
\usepackage[usenames,dvipsnames]{xcolor}
\hypersetup{colorlinks=true, linkcolor=BrickRed, urlcolor=blue!50!black, citecolor=blue!50!black}

\usepackage[normalem]{ulem}

\newcommand\diff{\mathrm{d}}
\renewcommand{\vec}[1]{\mathbf{#1}}
\renewcommand{\imath}[0]{\mathsf{i}}

\def\bra#1{\mathinner{\langle{#1}|}}
\def\ket#1{\mathinner{|{#1}\rangle}}
\def\braket#1{\mathinner{\langle{#1}\rangle}}

\usepackage[nodayofweek,level]{datetime}

\begin{document}

\title{Intermediate scattering function of an anisotropic Brownian circle swimmer}

\author{Christina Kurzthaler$^1$ and Thomas Franosch$^{1}$}

\affiliation{$^1$Institut f\"ur Theoretische Physik, Universit\"at Innsbruck, Technikerstra{\ss}e 21A,
A-6020 Innsbruck, Austria}
\email{thomas.franosch@uibk.ac.at}

\begin{abstract}
Microswimmers exhibit noisy
circular motion due to asymmetric propulsion mechanisms, their chiral body shape, or by hydrodynamic couplings in the vicinity of surfaces.
Here, we employ the Brownian circle swimmer model and  characterize theoretically the dynamics in terms of the directly measurable intermediate scattering function.
We derive the associated Fokker-Planck equation for the conditional probabilities and provide an exact solution in terms of generalizations of the Mathieu functions.
Different spatiotemporal regimes are identified reflecting
the bare translational diffusion at large wavenumbers, the persistent circular
motion at intermediate wavenumbers and an enhanced effective diffusion at small
wavenumbers. In particular, the circular motion of the particle manifests itself in
characteristic oscillations at a plateau of the intermediate scattering function for wavenumbers
probing the radius.
\end{abstract}
\date{\formatdate{26}{8}{2017}}

\maketitle

\section{Introduction}
A plethora of active agents ranging from biological microswimmers to
artificially synthesized self-propelled particles  exhibit peculiar dynamical
behavior while moving in aqueous media 
where Brownian motion plays a pivotal role~\cite{Romanczuk:2012,Vicsek:2012, Marchetti:2013, Elgeti:2015,
Bechinger:2016}.
These active particles are intrinsically out of equilibrium
and their transport properties are highly sensitive to
their body shape, the symmetry of the propulsion mechanism, and also
interactions with interfaces, which all have been shown to induce a chiral swimming
pattern.
Examples of circular motion close to surfaces include
sperms~\cite{Woolley:2003, Riedel:2005,Bohmer:2005,Friedrich:2008}, and
bacteria~\cite{Berg:1990,DiLuzio:2005,Lauga:2006, Hill:2007,Li:2008,DiLeonardo:2011,Utada:2014},
whereas a special type of algae, \textit{Chlamydomonas reinhardtii}, exhibits a
helical swimming trajectory due to an asymmetry in its flagella
beat~\cite{Racey:1981,Racey:1983,Martinez:2012}. As a consequence of an either
simple asymmetric shape or two internal motors propelling into different
directions, also artificial microswimmers such as asymmetric Janus
particles~\cite{Kummel:2013,tenHagen:2014}, bimetallic
micromotors~\cite{Bidoz:2005,Marine:2013, Takagi:2013}, and self-assembled
doublets of spherical Janus particles~\cite{Ebbens:2010} display circular
motion. Moreover, particles, that perform chemotaxis along their self-generated gradient,
are expected to follow circular trajectories in a strong chemical field~\cite{Taktikos:2011}.

As observed in experiments, a minor imbalance in the swimming motion can lead
to a rich dynamical behavior of these active agents on the macroscopic level.
Hence, a profound knowledge on different levels of coarse graining is necessary
to fully understand the non-equilibrium physics of these circle swimmers.
Hydrodynamic models characterizing the motion of a linked-bead swimmer in
bulk~\cite{Dreyfus:EPJ:2005,Ledesma:2012} and close to
walls~\cite{Dunstan:2012}, and also flagellated microswimmers at a
surface~\cite{Lauga:2006,Shum:2010,Hu:2015} including the full hydrodynamic
interactions have predicted circular swimming patterns of these active agents
using analytic computations and computer simulations.

In addition to these microscopic theories, mesoscopic models ignoring the
origin of the swimming motion and neglecting hydrodynamic interactions have
been elaborated in terms of effective non-equilibrium Langevin
equations~\cite{vanTeffelen:2008,
vanTeffelen2009,Ebbens:2010,Mijalkov:2013,Krueger:2016,Jahanshahi:2017}.  Here,
the main quantity of interest constitutes the mean-square
displacement~\cite{vanTeffelen:2008, vanTeffelen2009}, which exhibits an
intermediate oscillatory behavior as a genuine fingerprint of the circular
motion. These transport properties have been also quantified in experiments
using particle tracking and compared to analytical
predictions~\cite{Ebbens:2010,Utada:2014,Krueger:2016}.

Another experimentally accessible quantity that contains much more general spatiotemporal
information on the particle's dynamics constitutes the intermediate scattering function $F(k,t)$~\cite{Berne:1976,Dhont:1996}, which
measures the dynamics at lag time $t$ and length scales
$2\pi/k$. Mathematically, it is obtained by a Fourier transform of the
probability density of the displamcents and represents the associated
characteristic function~\cite{Gardiner:2009}.

Only recently the intermediate scattering function
has been computed analytically for simple run-and-tumble particles~\cite{Martens:2012} and for active Brownian
agents~\cite{Kurzthaler:2016}.
Whereas it has also been measured for \textit{Chlamydomonas reinhardtii}  by light
scattering experiments~\cite{Racey:1981,Racey:1983}, and within the recently
developed image based framework of differential dynamic
microscopy~\cite{Martinez:2012}, only approximations of the intermediate
scattering function of circle swimmers valid at rather small
length scales have been worked out~\cite{Racey:1983,Martinez:2012}.

Here, we derive the Fokker-Planck equation for the conditional probability density of the displacements of a
Brownian circle swimmer. These active agents display persistent circular motion, but are also subject to rotational
and anisotropic translational diffusion, which entails the rotational-translational
coupling. To quantify the dynamics of these particles, we provide an analytical solution
of the intermediate scattering function in terms of generalizations of the Mathieu functions. We numerically evaluate the intermediate
scattering function for the full range of length scales and identify different spatiotemporal regimes
reflecting the bare translational diffusion, the persistent circular motion and also the enhanced effective diffusion of the
circle swimmer.
Furthermore, we obtain the low-order moments such as the mean-square and mean-quartic displacement upon expansion of the intermediate
scattering function in the wavenumber.  In particular, we discuss the
non-Gaussian parameter and corroborate our results with computer simulations.

\section{The model}
\begin{figure}[htp]
  \includegraphics[width=\linewidth,keepaspectratio]{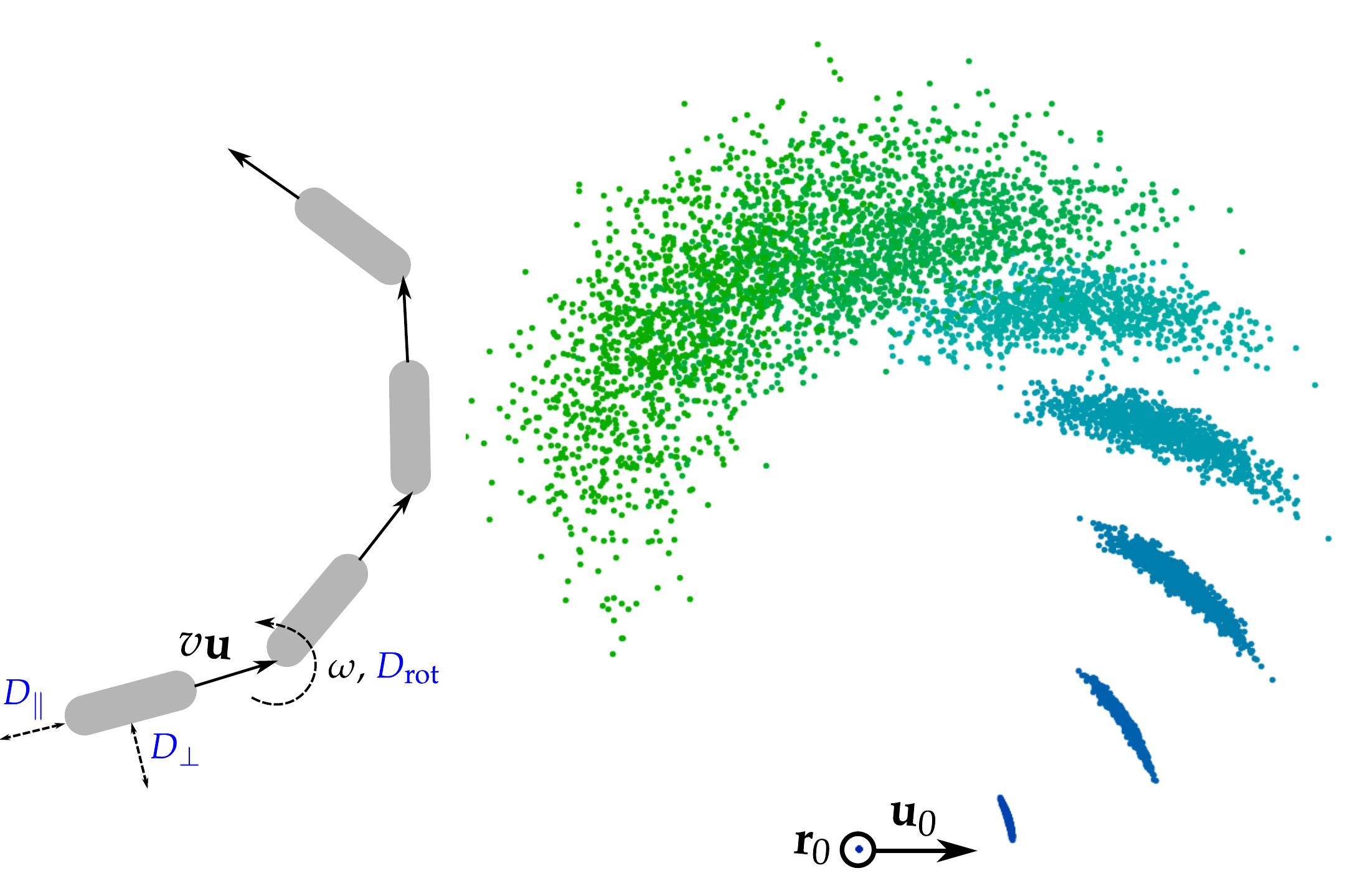}
  \caption{(Left) Model set up and (right) distribution of
anisotropic Brownian circle swimmers
with the same initial orientation $\vec{u}_0$, initial position $\vec{r}_0$,
P{\'e}clet number $\text{Pe}=\infty$, and quality factor $M=16$.
Here, different colors indicate subsequent time steps $t_i/\tau_\text{rot}\in[0, 0.04]$, where $\tau_\text{rot}$ denotes
the rotational diffusion time. \label{fig:dist}}
\end{figure}

We assume the particle to move in a plane with constant speed~$v$ along its instantaneous
orientation $\vec{u}(t)=[\cos\vartheta(t),\sin\vartheta(t)]^T$ parametrized by
the polar angle $\vartheta(t)$. To model the circular motion of the
active particle, the orientation $\vartheta(t)$ displays an average drift
of constant angular velocity $\omega$ and is also subject to orientational Brownian motion
characterized by the rotational diffusion coefficient $D_\text{rot}$.
Furthermore, the motion displays translational diffusion encoded in
the diffusion coefficients parallel $D_\parallel$ and perpendicular $D_\perp$
to the orientation, see Fig.~\ref{fig:dist}. Hence, the Langevin equations in It$\bar{\text{o}}$
form for the position $\vec{r}(t)$ and the orientation $\vartheta(t)$ of an
anisotropic circle swimmer assume the form~\cite{vanTeffelen:2008}:
\begin{align}
  \diff \vartheta   &= \omega  \diff t +\sqrt{2D_\text{rot}} \diff \psi,\label{eq:theta}\\
  \diff \vec{r}     &= v\vec{u}\diff t+ \left[\sqrt{2 D_\parallel}\vec{u}\vec{u}^T+\sqrt{2D_\perp}(\mathbb{I}-\vec{u}\vec{u}^T)\right]\diff \boldsymbol{\xi}.\label{eq:pos}
\end{align}
Here, the rotational and translational diffusion are modeled in terms of
independent Gaussian white noise processes $\psi(t)$ and $\boldsymbol{\xi}(t)$ of zero
mean and delta correlated variance $\langle\psi(t)\psi(t')\rangle = \delta(t-t')$ and
$\langle\xi_i(t)\xi_j(t')\rangle = \delta_{ij}\delta(t-t')$ for $i,j=1,2$,
respectively.
Although there is multiplicative noise in the translational motion, it does not induce
noise induced drift since the noise only couples to the orientation. As a consequence the equations assume the same form
also in the Stratonovich interpretation.
Typical distributions of circle swimmers,
obtained from simulations of the Langevin equations
(see Appendix~\ref{ap:simulation}), reveal a narrow distribution at short times,
which displays a circle and thereby broadens due to rotational diffusion at longer times [Fig.~\ref{fig:dist}].

\begin{figure}[htp]
  \includegraphics[width=\linewidth,keepaspectratio]{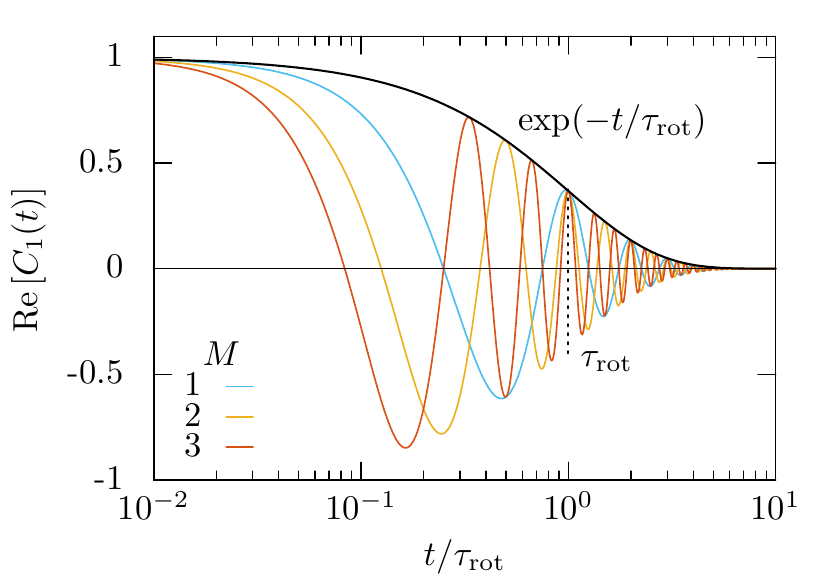}
  \caption{Real part of the angular correlation function $C_1(t)$ for different quality factors $M$.\label{fig:angular}}
\end{figure}

To quantify the deterministic circular motion with respect to
the rotational diffusion we introduce the dimensionless quality factor $M$
\begin{align} M &= \frac{\omega/2\pi}{D_\text{rot}}=\frac{\omega}{2\pi}\tau_\text{rot},
\end{align}
where $\tau_\text{rot}:= 1/D_\text{rot}$ denotes the rotational diffusion time.
Then the angular correlation function of order $n$,
$C_n(t) = \langle\exp\left[\imath n \left(\vartheta(t)-\vartheta(0)\right)\right]\rangle$
fulfills the equation of motion (see Appendix~\ref{ap:M} for a derivation)
\begin{align}
 \frac{\diff}{\diff t}C_n(t) -nD_\text{rot}(2\pi\imath M-n)C_n(t)&=0.
\end{align}
The solution
\begin{align}
C_n(t)&=\exp(-n^2t/\tau_\text{rot})\exp(2\pi \imath n Mt/\tau_\text{rot}),
\end{align}
becomes complex, which is a fingerprint of a non-equilibrium process. Here, $\exp(-n^2t/\tau_\text{rot})$ constitutes the envelope of the oscillations of
the real and imaginary part of the angular correlation function (see Fig.~\ref{fig:angular}).
In particular, we observe that the quality factor $M$ measures the number of
circles a swimmer has completed within the rotational diffusion time $\tau_\text{rot}$.

We obtain another dimensionless quantity $\Delta D/\bar{D}$ as ratio between the translational anisotropy
$\Delta D=D_\parallel-D_\perp$ and the mean translational diffusion coefficient
$\bar{D}=(D_\parallel+D_\perp)/2$.  To illustrate our results we follow Ref.~\cite{vanTeffelen:2008} and use the anisotropy
$D_\parallel=2D_\perp$ inspired by the hydrodynamics for passive
rod-like particles in the limit of very large aspect ratio~\cite{Doi:1986}.
In addition, we also define the P{\'e}clet
number by $\text{Pe}=av/\bar{D}$, where $a$ denotes
the characteristic length  $a=\sqrt{3\bar{D}/D_\text{rot}}/2$. For a spherical particle in equilibrium $a$
corresponds merely to the radius of the particle. The P{\'e}clet number quantifies the relative importance of the persistent swimming motion
of the particle with respect to translational diffusion.

The fundamental quantity of interest constitutes the experimentally
measurable intermediate scattering function (ISF)~\cite{Berne:1976, Martinez:2012}
\begin{align}
  F(\vec{k},t)&= \langle\exp[-\imath\vec{k}\cdot\Delta\vec{r}(t)]\rangle.\label{eq:characteristic}
\end{align}
The ISF can be interpreted as the characteristic function of the random
displacement variable $\Delta\vec{r}(t)$ and the moments of the stochastic process
are obtained as derivatives of the ISF with respect to the
wavenumber~\cite{Gardiner:2009}.  The ISF of the circle swimmer can be computed by
\begin{align}
  F(\vec{k},t)&= \int_0^{2\pi} \diff\vartheta \int_0^{2\pi}\frac{\diff\vartheta_0}{2\pi} \widetilde{\mathbb{P}}(\vec{k},\vartheta,t|\vartheta_0),\label{eq:ISF}
\end{align}
where $\widetilde{\mathbb{P}}(\vec{k},\vartheta,t|\vartheta_0)$ denotes the spatial Fourier transform
 \begin{align}
  \widetilde{\mathbb{P}}(\vec{k},\vartheta,t|\vartheta_0)&= \int_{\mathbb{R}^2} \diff^2 r \exp(-\imath \vec{k}\cdot\vec{r})\mathbb{P}(\vec{r},\vartheta,t|\vartheta_0),
\end{align}
of the conditional probability density
$\mathbb{P}(\Delta\vec{r}=\vec{r}-\vec{r}_0,\vartheta,t|\vartheta_0)$ to find a particle at position $\vec{r}$ with orientation $\vartheta$ after a lag time $t$
given that it has been at $\vec{r}_0$ with  orientation $\vartheta_0$ at zero lag time, $t=0$.

After averaging over the orientations [Eq.~\eqref{eq:ISF}], the motion of the particle is isotropic, and therefore the
ISF evaluates to a real function depending on the magnitude of the wavevector $k=|\vec{k}| $ only,
$F(\vec{k},t)=F(k,t)$. In particular, after averaging Eq.~\eqref{eq:characteristic} over the directions $\varphi = \angle(\vec{k},\Delta\vec{r}(t))$ of the wavevector $\vec{k}$ the
ISF reduces to the Bessel function of order zero~\cite{Arfken:2005}
\begin{align}
  F(k,t)  &= \left\langle\int_0^{2\pi}\frac{\diff\varphi}{2\pi}\exp\left(-\imath k|\Delta\vec{r}(t)|\cos\varphi\right)\right\rangle\\
  &=\langle J_0(k|\Delta\vec{r}(t)|)\rangle.\label{eq:Bessel}
\end{align}
To obtain an analytic expression for the ISF, we start from Eq.~\eqref{eq:ISF} and compute the Fourier transform of
the probability density. Therefore we first derive the
Fokker-Planck equation for the probability density
$\mathbb{P}\equiv\mathbb{P}(\Delta\vec{r},\vartheta,t|\vartheta_0)$, which is an equivalent
description of the motion of the circle swimmer as the Langevin equations
[Eq.~\eqref{eq:theta}-\eqref{eq:pos}].  We obtain by standard methods of
stochastic calculus~\cite{Gardiner:2009} the equation of motion
\begin{align}
  \partial_t \mathbb{P} &= -v\vec{u}\cdot\partial_\vec{r}\mathbb{P}-\omega\partial_\vartheta\mathbb{P}+\partial_\vec{r}\cdot (\vec{D}\cdot\partial_\vec{r}\mathbb{P})+D_\text{rot}\partial_\vartheta^2\mathbb{P},\label{eq:fp}
\end{align}
where $\vec{D}$ denotes the translational diffusion tensor $\vec{D}= D_\perp
\mathbb{I}+\Delta D \vec{u}\vec{u}^T$, which couples the translational
diffusion to the orientation of the particle.  The first two advective terms on
the right-hand side of Eq.~\eqref{eq:fp} describe the deterministic active motion and
rotational drift of the particle.  The remaining terms encode the translational and
rotational diffusion, respectively.  Then the equation of motion for
the corresponding Fourier transform follows
\begin{align}
  \partial_t \widetilde{\mathbb{P}}&= -\imath v\vec{u}\cdot\vec{k}\widetilde{\mathbb{P}}-\omega\partial_\vartheta \widetilde{\mathbb{P}}- [D_\perp \vec{k}^2+\Delta D(\vec{u}\cdot\vec{k})^2]\widetilde{\mathbb{P}}\notag\\
                                    &\qquad +D_\text{rot}\partial^2_\vartheta\widetilde{\mathbb{P}}.\label{eq:fourier}
\end{align}
Counterclockwise swimmers are related to clockwise swimmers by mirror symmetry. In fact,
Eq.~\eqref{eq:fourier} remains invariant under simultaneous reflections of
the wavevector $\vec{k} \mapsto \vec{k}'$ and the orientation $\vec{u} \mapsto \vec{u}'$ across an arbitrarys axis,
provided the angular velocity changes sign $\omega\mapsto -\omega$.
Choosing as axis of reflection the direction of $\vec{k}$ reveals that
the ISF is insensitive to the chirality of the particle and does not allow
distinguishing whether the agent displays a clockwise or counterclockwise circular swimming motion.

Special cases of the previous equation [Eq.~\eqref{eq:fourier}] have already been solved in terms of Mathieu functions for a
passive anisotropic Brownian particle ($v=0$ and $\omega=0$)~\cite{Munk:2009},
and also for a three dimensional anisotropic passive~\cite{Leitmann:2016} and active Brownian particle ($\omega=0$)~\cite{Kurzthaler:2016}.
Yet, no solution for the ISF of a Brownian circle swimmer has been elaborated up to now.

We solve Eq.~\eqref{eq:fourier} by an expansion of $\widetilde{\mathbb{P}}$ in terms of appropriate eigenfunctions.
We choose the direction of the wavevector $\vec{k}=k\vec{e}_x$
such that the equation of motion [Eq.~\eqref{eq:fourier}] reduces to
\begin{align}
  \partial_t \widetilde{\mathbb{P}}&= \bigl[-\imath vk\cos\vartheta-\omega\partial_\vartheta - \left(D_\perp +\Delta D\cos^2\vartheta\right)k^2\notag\\
  & \qquad
                                    +D_\text{rot}\partial^2_\vartheta\bigr]\widetilde{\mathbb{P}}.\label{eq:fourier2}
\end{align}
Hence, separation of variables $\exp(-\lambda t)z(\vartheta)$ in terms of angular eigenfunctions $z(\vartheta)$ yields the eigenvalue problem
\begin{align}
\Bigl(\frac{\diff^2}{\diff\vartheta^2}-\frac{\imath vk}{D_\text{rot}}\cos\vartheta-\frac{\Delta D k^2}{D_\text{rot}} \cos^2\vartheta-2\pi M\frac{\diff}{\diff\vartheta}\notag\\
 -\frac{D_\perp k^2}{D_\text{rot}}+\frac{\lambda}{D_\text{rot}}\Bigr)z(\vartheta) =0,\label{eq:eigenvalue}
\end{align}
reminiscent of the Mathieu equation~\cite{NIST:print,NIST:online}. To make connection with the standard form of the Mathieu
equation we use a change of variables $x=\vartheta/2$ and obtain
an eigenvalue problem $L(q,c,M) z(x) = a z(x)$  for the
non-hermitian Sturm-Liouville operator
\begin{align}
L&=L(q,c,M)\notag\\
&= -\frac{\diff^2}{\diff x^2}+2q\cos(2x)+c^2\cos^2(2x)+4\pi M\frac{\diff}{\diff x},\label{eq:Sturm}
\end{align}
dependent on the dimensionless parameters $q,c$, and quality factor $M$.
Here, we identify the  deformation parameters $q = 2\imath vk/D_\text{rot}$ and $c^2 = 4\Delta D k^2/D_\text{rot}$ similar to the  generalized spheroidal wave functions in Ref.~\cite{Kurzthaler:2016}. In our case, $q$ is purely imaginary, and $c^2$ is real but may assume both signs.
The separation constant $\lambda$ is connected to the eigenvalue $a$ by $\lambda = D_\perp k^2+aD_\text{rot}/4$.

The eigenfunctions $z(q,c,M,x)$ and the corresponding  eigenvalues $a=a(q,c,M)$ are in general complex.
Since $q$ is imaginary the Sturm-Liouville operator displays a symmetry: if $z(q,c,M,x)$ is an eigenfunction of  $L(q,c,M)$ with eigenvalue $a(q,c,M)$, then $z(q,c,M,x+\pi/2)^*$ is eigenfunction to eigenvalue $a(q,c,M)^*$. Therefore complex eigenvalues come in complex conjugate pairs.  Furthermore, $z(q,c,M,-x)$ is eigenfunction to $L(q,c,-M)$ with eigenvalue $a(q,c,M)$, thus the spectrum does not depend on the chirality of the swimmer.

For the case that $c=M=0$ one recovers the Mathieu equation~\cite{NIST:print,NIST:online}, and by the change of variables we need only the $\pi$-periodic even and odd eigenfunctions $\text{ce}_{2n}(q,x)$ and $\text{se}_{2n+2}(q,x)$. These Mathieu functions are essentially deformed sines and cosines. For $c\neq 0$ but $M=0$
the eigenfunctions remain even and odd functions, and therefore are merely deformations of the Mathieu functions.
However, for circle swimmers $M\neq0$ the operator [Eq.~\eqref{eq:Sturm}] is no longer invariant under a parity transformation $x\mapsto -x$,
and the corresponding eigenfunctions are neither odd nor even in $x$.  Therefore, rather than deforming sines and cosines
we rely from the very beginning on deformations of the complex exponentials $\exp(2 n \imath x)$:
\begin{align}
  \text{ee}_{2n}(q,c,M,x) &= \sum_{m=-\infty}^\infty A_{2m}^{2n} e^{ 2m \imath x},\label{eq:fm}
\end{align}
where $A_{2m}^{2n}$ denotes the $2m-$th Fourier coefficient of $\text{ee}_{2n}(q,c,M,x)$.
The  eigenfunctions of $L(q,c,M)$ are therefore generalizations of the Mathieu functions $\text{ee}_{2n}(q,c,M,x)$, $n\in\mathbb{Z}$.
In Appendix~\ref{appendix:nonhermitian} we show that these eigenfunctions are orthogonal in the sense of
\begin{equation}
\int_0^{\pi} \! \diff x \ \text{ee}_{2m}(q,c,M,x)\text{ee}_{2n}(q,c,M,-x)=\pi\delta_{mn} . \label{eq:norm}
\end{equation}
Thus, the general solution of the Fourier transform $\widetilde{\mathbb{P}}$ is found in terms of an expansion of the corresponding eigenfunctions
\begin{align}
\lefteqn{\widetilde{\mathbb{P}}(\vec{k},\vartheta,t|\vartheta_0)=
 \frac{e^{-D_\perp k^2 t}}{2\pi}\sum_{n=-\infty}^\infty e^{-a_{2n} D_\text{rot} t/4}}\notag\\
 &\qquad\qquad\times\text{ee}_{2n}(q,c,M,\vartheta/2)\text{ee}_{2n}(q,c,M,-\vartheta_0/2).
\end{align}
By completeness of the eigenfunctions, this reproduces indeed the initial condition $\delta(\vartheta-\vartheta_0 \text{ mod }2\pi)$
for $t=0$.  
Performing the integrals in Eq.~\eqref{eq:ISF},  we obtain the analytic
expression of the ISF, which constitutes the principal result of this work
\begin{align}
\lefteqn{F(k,t)  =  \frac{e^{-D_\perp k^2 t}}{4\pi^2}\sum_{n=-\infty}^\infty e^{-a_{2n} D_\text{rot}t/4}}\notag\\
&\qquad\qquad \times\left[\int_0^{2\pi}\diff\vartheta \ \text{ee}_{2n}(q,c,M,\vartheta/2)\right]^2.\label{eq:ISFanalytics}
\end{align}
Here, the ISF depends explicitely on the diffusion coefficient perpendicular
to the particle's orientation, $D_\perp$, the anisotropy is hidden in the parameter $c$, which vanishes for
isotropic diffusion.
The ISF can then be efficiently evaluated numerically, see Appendix~\ref{numerics}.

\begin{figure*}[htp]
    \includegraphics[width=\linewidth,keepaspectratio]{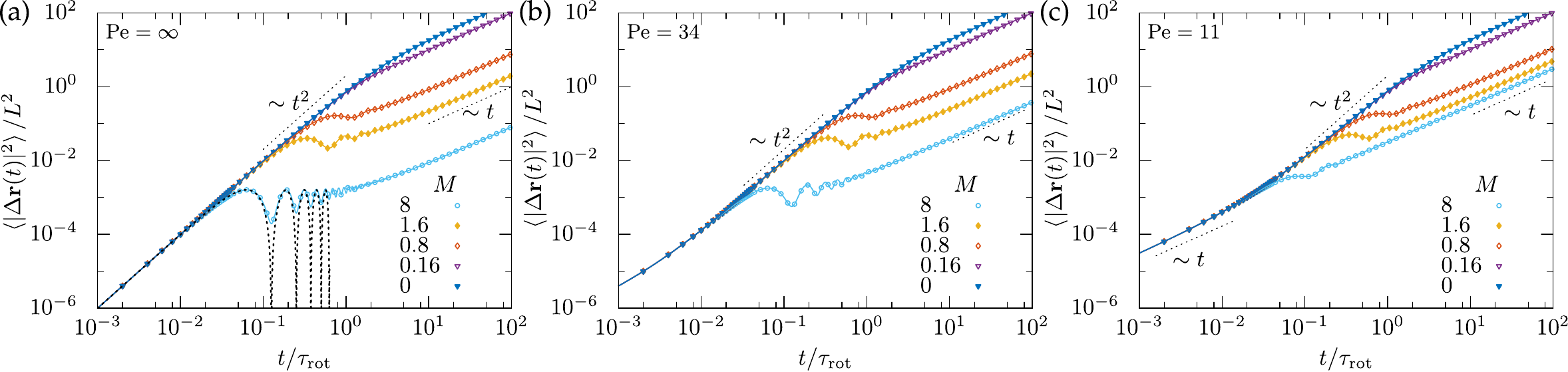}
    \caption{Mean-square displacement $\langle|\Delta\vec{r}(t)|^2\rangle/L^2$ in
      units of the persistence length $L=v/D_\text{rot}$ for an
      anisotropic ($\Delta D/\bar{D}=2/3$) Brownian circle swimmer for different P{\'e}clet numbers
      $\text{Pe}=av/\bar{D}$ and quality factors $M=\omega\tau_\text{rot}/2\pi$ with rotational diffusion time $\tau_\text{rot}=1/D_\text{rot}$.
       The black-dashed line in Fig. (a) indicates the mean-square displacement of a deterministic circle swimmer $\langle|\Delta\vec{r}(t)|^2\rangle = 4R^2\sin^2(\omega t/2)$,
      where $R=v/\omega$ is the radius of the circular motion. Simulation and theory results are shown with symbols and lines,
    respectively.\label{fig:msd}}
\end{figure*}
\section{Exact low-order moments\label{sec:perturbation}}
Most studies consider the low-order moments, such as the mean-square displacement, of active agents only.
Since the ISF can be viewed as the moment-generating function of the random displacements, the exact moments can
be obtained as a byproduct of our analysis. For consistency we evaluate the mean-square displacement, which has been calculated earlier~\cite{vanTeffelen:2008, Ebbens:2010},
and compute for the first time the mean-quartic displacement of an anisotropic Brownian circle swimmer.
These are then used as input for the non-Gaussian parameter.

To determine the exact low-order moments of the stochastic process, we expand  the
ISF [Eq.~\eqref{eq:Bessel}]  up to the fourth order in the wavenumber $k$
\begin{align}
  F(k,t)  &= 1-\frac{k^2}{4}\langle|\Delta\vec{r}(t)|^2\rangle +\frac{k^4}{64}\langle |\Delta\vec{r}(t)|^4\rangle +\mathcal{O}(k^6).\label{eq:sincex}
\end{align}
Therefore, we apply a time-dependent
perturbation theory for small wavenumbers $k$ in the form of a Dyson series~\cite{Sakurai:2011}. 
For convenience, we rely on the Dirac notation and introduce the scalar product
$\langle\varphi|\psi\rangle=(2\pi)^{-1}\int_0^{2\pi}\diff\vartheta\varphi^*(\vartheta)\psi(\vartheta)$.
Furthermore, we use the generalized angular basis
$\{\ket{\vartheta}\}$, which is orthogonal
$\braket{\vartheta|\vartheta_0}=2\pi \delta(\vartheta - \vartheta_0)$ and fulfills the closure relation
$(2\pi)^{-1}\int_{0}^{2\pi}\diff\vartheta|\vartheta\rangle\langle\vartheta| = \mathbb{I}$.  Then the  isomorphism between
angular functions $\psi(\vartheta)$ and states in the Hilbert space becomes manifest $|\psi \rangle \leftrightarrow \psi(\vartheta) = \langle \vartheta | \psi\rangle$.
Similarly, we define a time-evolution  operator $\widetilde{\mathbb{P}}(\vec{k},t)$ via its
matrix elements $\widetilde{\mathbb{P}}(\vec{k},\vartheta,t|\vartheta_0)=\langle\vartheta|\widetilde{\mathbb{P}}(\vec{k},t)|\vartheta_0\rangle/2\pi$,
such that $\widetilde{\mathbb{P}}(\vec{k},t=0)=\mathbb{I}$. 
Furthermore, we introduce a generator for the unperturbed time evolution $\hat{H}_0$  by $\langle
\vartheta | \hat{H}_0 | \psi \rangle = (\omega\partial_\vartheta- D_{\text{rot}}\partial_\vartheta^2) \psi(\vartheta)$ and a perturbation $\hat{V}$ via
$\langle \vartheta | \hat{V} | \psi \rangle
= \left(\imath v k \cos\vartheta + D_\perp k^2 + \Delta D k^2  \cos^2\vartheta \right) \psi(\vartheta)$.
We split the latter into two terms, $\langle\vartheta|\hat{V}_\text{swim}|\psi\rangle = \imath v k \cos\vartheta \psi(\vartheta)$ and
$\langle\vartheta|\hat{V}_\text{diff}|\psi\rangle = (D_\perp k^2 +\Delta D k^2 \cos^2\vartheta) \psi(\vartheta)$,
containing perturbations in first and second order in $k$, respectively.

With these definitions we rewrite Eq.~\eqref{eq:fourier} in operator form
\begin{align}
\partial_t\widetilde{\mathbb{P}}(\vec{k},t)&=-\hat{H}_0\widetilde{\mathbb{P}}(\vec{k},t)-\hat{V}\widetilde{\mathbb{P}}(\vec{k},t). \label{eq:Schroedinger}
\end{align}
The eigenfunctions of the unperturbed operator $\hat{H}_0$ are simply the Fourier modes $| n \rangle $ with angular representation
$\braket{\vartheta | n} =  \exp(\imath n \vartheta)$.
The corresponding unperturbed eigenvalues read $a_n^{0} = D_{\text{rot}} n^2  +\imath n \omega$, in particular, $\hat{H}_0
\ket{n} = a_n^0\ket{n}$. The basis fulfills the normalization condition $\braket{m|n}= \delta_{mn}$
and closure relation $\sum_{n=-\infty}^{\infty}\ket{n}\bra{n}= \mathbb{I}$.
The ISF can then be expressed in terms of the eigenbasis of the unperturbed operator by sandwiching the closure relation
\begin{align}
  F&(k,t)  = \int_0^{2\pi}\frac{\diff\vartheta}{2\pi}\int_0^{2\pi}\frac{\diff\vartheta_0}{2\pi}\langle\vartheta|\widetilde{\mathbb{P}}(\vec{k},t)|\vartheta_0\rangle \nonumber \\
   &= \sum_{m,n=-\infty}^\infty\int_0^{2\pi}\frac{\diff\vartheta}{2\pi}\int_0^{2\pi}\frac{\diff\vartheta_0}{2\pi}\langle\vartheta|m\rangle\langle m|\widetilde{\mathbb{P}}(\vec{k},t)|n\rangle \langle n|\vartheta_0\rangle \nonumber \\
   &= \sum_{m,n=-\infty}^\infty\int_0^{2\pi}\frac{\diff\vartheta}{2\pi} e^{\imath m\vartheta}\int_0^{2\pi}\frac{\diff\vartheta_0}{2\pi} e^{-\imath n \vartheta_0}\langle m|\widetilde{\mathbb{P}}(\vec{k},t)|n\rangle \nonumber \\
   &= \langle m=0|\widetilde{\mathbb{P}}(\vec{k},t)|n=0\rangle.\label{eq:ISFexpanded}
\end{align}
To obtain the matrix elements of $\widetilde{\mathbb{P}}(\vec{k},t)$, we solve Eq.~\eqref{eq:Schroedinger} in
terms of a  Dyson series \cite{Sakurai:2011}
\begin{align}
  \widetilde{\mathbb{P}}(\vec{k},t) &= e^{-\hat{H}_0  t}-\int_0^t\diff s \ \ e^{-\hat{H}_0(t-s)}\hat{V} e^{-\hat{H}_0s}+\mathcal{O}(\hat{V}^2). ~\label{eq:dyson}
\end{align}
The initial condition of the time-evolution operator is
$\widetilde{\mathbb{P}}(\vec{k},t=0)=\mathbbm{1}$, and the unperturbed solution of
the propagator is formally expressed by $e^{-\hat{H}_0t}$.
Hence, the matrix elements of the time-evolution operator up to first order in the perturbation $\hat{V}$ are computed by
\begin{align}
\langle m |\widetilde{\mathbb{P}}(\vec{k},t)|n \rangle &= e^{-a_m^0 t}\left[\delta_{mn}-\langle m| \hat{V}|n\rangle \int_0^t \diff s \ e^{-(a_n^0-a_m^0)s}\right],
\end{align}
with the matrix elements of the perturbation
\begin{align}
  \langle m |\hat{V}|n\rangle &= \left(D_\perp+\frac{\Delta D}{2}\right)k^2\delta_{mn}+\frac{\imath vk}{2}\left(\delta_{m,n+1}+\delta_{m,n-1}\right)\notag\\
                              & \ \ \ \ +\frac{\Delta Dk^2}{4}\left(\delta_{m,n+2}+\delta_{m,n-2}\right).\label{eq:vmn}
\end{align}
To obtain an expansion of the ISF up to the fourth order in the wavenumber
$\mathcal{O}(k^4)$ we have to extend the Dyson series up to the order
$\mathcal{O}(\hat{V}_\text{swim}^4,\hat{V}_\text{diff}^2,
\hat{V}_\text{swim}^2\hat{V}_\text{diff},\hat{V}_\text{diff}\hat{V}_\text{swim}^2,\hat{V}_\text{swim}\hat{V}_\text{diff}\hat{V}_\text{swim})$. Then the zeroth matrix element
$m=n=0$ of the Fourier transform constitutes the expanded ISF up to $k^4$.  The
computations are rather lengthy but analog to the ones, that have been
presented before.

\subsection{Mean-square displacement}
Evaluating the expansion of the ISF for small wavenumbers [Eq.~\eqref{eq:ISFexpanded}]
and comparing with Eq.~\eqref{eq:sincex}, we
obtain the mean-square displacement of an anisotropic Brownian circle swimmer
\begin{align}
  \langle|\Delta\vec{r}(t)|^2\rangle &= \frac{2e^{-D_\text{rot}t}v^2}{\left(D_\text{rot}^2+\omega ^2\right)^2}\Bigl[e^{D_\text{rot} t} \bigl(D_\text{rot}^2 (D_\text{rot} t-1)\notag\\
   & \ \ +\omega ^2 (D_\text{rot} t+1)\bigr)
  +(D_\text{rot}^2-\omega^2)\cos(\omega t)\notag\\
  &\ \ -2 D_\text{rot} \omega  \sin (\omega t)\Bigr]+4\bar{D}t,\label{eq:msd}
\end{align}
which has already been computed in Ref.~\cite{vanTeffelen:2008, Ebbens:2010}.

The problem displays three characteristic times:
the time the particle diffuses translationally before active motion dominates,
$\tau_\text{diff}:= \bar{D}/v^2$, the rotational diffusion time $\tau_\text{rot}$ and
the time a particle needs to complete a circle, $\tau_\omega:= 2\pi/\omega$.
If $\tau_\omega,\tau_\text{rot}\lesssim\tau_\text{diff}$ translational diffusion
dominates the dynamics of the active particle for all times. 
Similarly, if $\tau_\omega \gtrsim \tau_\text{rot}$, the orientation of the particle becomes randomized
before the particle can even complete a circle. Therefore, only the ordering $\tau_\text{diff}\lesssim\tau_\omega \lesssim\tau_\text{rot}$
displays novel physics and we restrict the discussion to this case.

For finite P{\'e}clet numbers the mean-square displacement increases linearly
with diffusion coefficient $\bar{D}$ at times $t\lesssim \tau_\text{diff} $, see Fig.~\ref{fig:msd} (b) and (c).
Then, for intermediate times
the mean-square displacement increases quadratically in time $v^2 t^2$ due to the
persistent swimming motion. Within the rotational diffusion time $\tau_\text{rot}$
a particle without torque $M=0$ covers a typical distance $L=v/D_\text{rot}$, denoted by the persistence length.

\begin{figure*}[htp]
  \includegraphics[width=\linewidth,keepaspectratio]{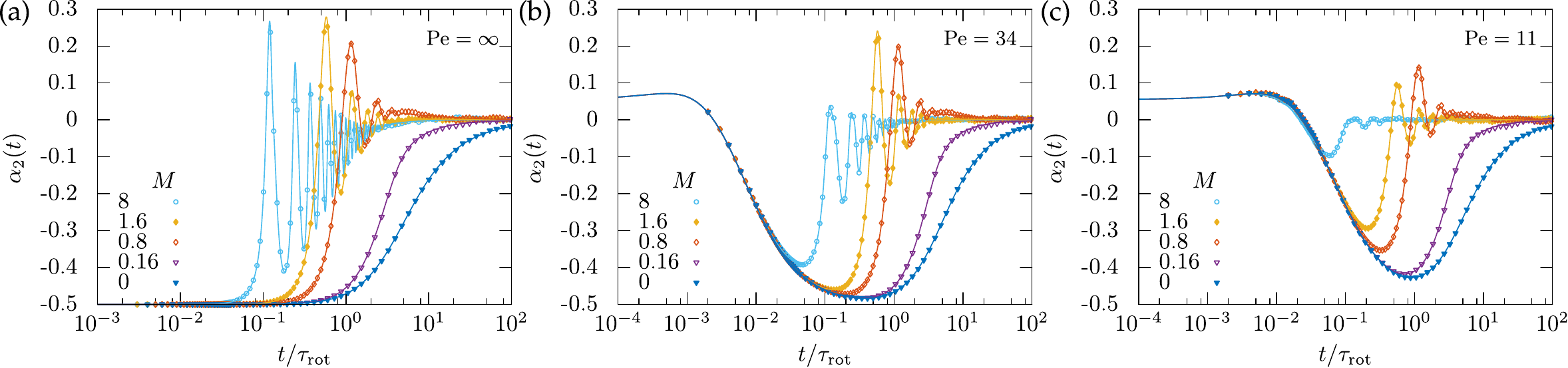}
  \caption{Non-Gaussian parameter $\alpha_2(t)$ for an anisotropic Brownian
  circle swimmer with hydrodynamic anisotropy $\Delta D/\bar{D}=2/3$ for
different P{\'e}clet numbers $\text{Pe}=av/\bar{D}$ and quality factors
$M=\omega\tau_\text{rot}/2\pi$  with rotational diffusion time $\tau_\text{rot}=1/D_\text{rot}$. Simulation and theory results are shown with
symbols and lines, respectively.\label{fig:ngp}}
\end{figure*}

For Brownian circle swimmers the ballistic increase is followed by oscillations
at times $t \gtrsim \tau_\omega$, where the particle has completed a circle.
In the case of a deterministic swimmer, $D_\text{rot}=\bar{D}=0$,
the displacement due to the persistent circular motion is thus
$|\Delta\vec{r}(t)|=2R|\sin(\omega t/2)|$, 
where $R$ denotes the radius of the circle, $R:=v/\omega$. Then the moments
of the displacements evaluate to $\langle|\Delta\vec{r}(t)|^n\rangle = \left|2R\sin(\omega t/2)\right|^n$,
which is also indicated in Fig.~\ref{fig:msd} (a).
Hence, the oscillations can be rationalized by the circular motion only, whereas
the fading out of the oscillations at longer times is due to the rotational diffusion.
In particular, the number of circles $M$ the particle swims during the rotational diffusion time is reflected in
the number of oscillations in the mean-square displacement. These oscillations
are also smeared out due to translational diffusion for decreasing P{\'e}clet numbers
at short times [Fig.~\ref{fig:msd}].

For long times the mean-square displacement evolves again
linearly in time and we obtain an effective diffusion coefficient
$D_\text{eff}= \bar{D}+v^2D_\text{rot}/[2(D_\text{rot}^2+\omega^2)]$.  In
particular, we observe an enhancement of the bare diffusion which is reduced by
the torque $D_\text{eff}/\bar{D}=1+2 \ \text{Pe}^2/[3(1+4\pi^2 M^2)]$.

\subsection{Non-Gaussian parameter}
A sensitive indicator that measures the deviation of the stochastic process
from a Gaussian process constitutes the non-Gaussian parameter~\cite{Hofling:2013}, which is
defined in 2D by
\begin{align}
\alpha_2(t)&= \frac{1}{2}\frac{\langle|\Delta\vec{r}(t)|^4\rangle}{\langle|\Delta\vec{r}(t)|^2\rangle^2}-1.
\end{align}
The expression of the mean-quartic displacement is rather lengthy and we refer
to Appendix~\ref{appendix:meanquartic} [Eq.~\eqref{eq:mqd}].

For long times $t\gtrsim\tau_\text{rot}$ the non-Gaussian parameter tends towards
zero for all P{\'e}clet numbers and angular velocities, since the motion gets
randomized and evolves to an effective diffusion, see Fig.~\ref{fig:ngp}.

For infinite P{\'e}clet number $\text{Pe}=\infty$ we observe a negative
non-Gaussian parameter at short times, $\alpha_2(t\rightarrow 0)
\rightarrow-1/2$ independent of the angular velocity $\omega$, whereas for
finite P{\'e}clet number the non-Gaussian parameter approaches a non-negative
constant $\alpha_2(t\rightarrow 0)\rightarrow \Delta D^2/8\bar{D}^2$.  Note,
that the non-Gaussian parameter vanishes for isotropic particles at short
times ($\Delta D = 0$).

In the case of a deterministic circle swimmer, $D_\text{rot}=\bar{D}= 0$, we find
that the non-Gaussian parameter evaluates to a constant $\alpha_2(t)=-1/2$ for all times.
This value is indeed observed in the full solution for $\text{Pe}=\infty$ and
times $t\lesssim \tau_\text{rot}$ where the rotational diffusion
has not yet set in (see Fig.~\ref{fig:ngp} (a)).

At intermediate times the non-Gaussian parameter for finite quality factors $M$
approaches the non-Gaussian parameter of an active Brownian particle
with $M=0$ reflecting the active swimming motion.  This regime is followed by an
oscillatory behavior that can be rationalized by the interplay of deterministic circular
motion and rotational diffusion of the particle. In particular, similar to the mean-square displacement
[Fig.~\ref{fig:msd}] the number $M$ of oscillations for $t\lesssim\tau_\text{rot}$
reflects the number of fulfilled circles within the rotational diffusion time.
However, these oscillations smear out for finite P{\'e}clet number at short times,
due to the translational diffusion.

\begin{figure*}[htp]
  \includegraphics[width=\linewidth,keepaspectratio]{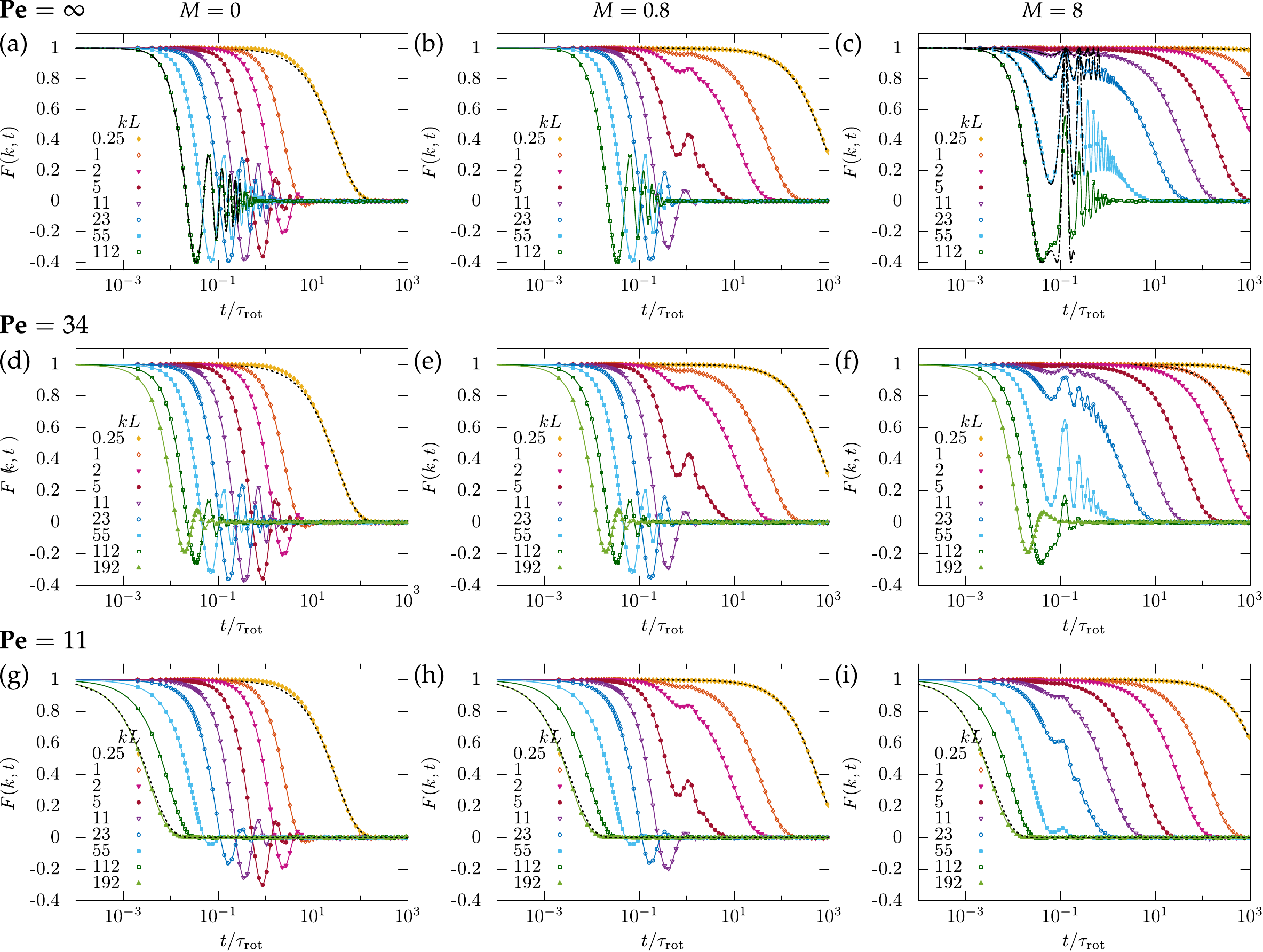}
  \caption{Intermediate scattering function $F(k,t)$ for a single, anisotropic
  circle swimmer with hydrodynamic anisotropy $\Delta D/\bar{D}=2/3$ for
different P{\'e}clet numbers $\text{Pe}=av/\bar{D}$ and qualitiy factors
$M=\omega\tau_\text{rot}/2\pi$. The dashed line represents the effective diffusion $\exp(-D_\text{eff}k^2t)$
and bare diffusion $\exp(-\bar{D}k^2t)$ for large and small
wavenumbers, respectively. The dashed-dotted line in Fig.~(a) indicates the
Bessel function of order zero, $J_0(vkt)$, and in Fig.~(c) the ISF for persistent circular motion, $J_0(2kR|\sin\omega t /2|)$. Simulation and theory results are shown with symbols and
lines, respectively.\label{fig:ISF}}
\end{figure*}

\section{Intermediate scattering function}
We have evaluated numerically the intermediate scattering function
[Eq.~\eqref{eq:ISFanalytics}] for arbitrary times and a wide range of length
scales measured in terms of the persistence length $kL$, and compare different
P{\'e}clet numbers $\text{Pe}$ and quality factors $M$, see Fig.~\ref{fig:ISF}.

For small wavenumbers the ISF can be approximated
by an enhanced effective diffusion $\exp(-D_\text{eff}k^2t)$, where the
diffusion coefficient $D_\text{eff}$ is taken from the slope of the mean-square
displacement at long times [Eq.~\eqref{eq:msd}]. In particular, a reduction of
the effective diffusion is observed with increasing quality factor $M$
[Fig.~\ref{fig:ISF}].

For large wavenumbers and P{\'e}clet number $\text{Pe}=11$ [Fig.~\ref{fig:ISF}
~(g)-(i)] the intermediate scattering function again approaches an
exponential $\exp(-\bar{D}k^2t)$ reflecting the bare translational diffusion.
A similar behavior occurs for $\text{Pe}=34$ at even higher wavenumbers (not shown).

In contrast, for infinite P{\'e}clet number
$\text{Pe}=\infty$ [Fig.~\ref{fig:ISF}~(a)-(c)] and vanishing quality factor $M=0$ the trajectories can be
approximated by a pure persistent motion  $|\Delta \vec{r}(t)|=vt$, in
particular, the ISF then assumes the form $F(k,t)=J_0(vkt)$, as
indicated by the dashed-dotted line in Fig.~\ref{fig:ISF}~(a). Note that the
approximation is only illustrated for $t/\tau_\text{rot}\lesssim 0.3$, as for
longer times 
rotational diffusion washes out the oscillations of the Bessel function.

The circular motion of the particle ($M\neq0$) first becomes apparent in the
ISF at times $t\sim\tau_\omega$ [Fig.~\ref{fig:ISF}~(b),(c),(e),(f),(h),(i)],
where the particle completes a full circle of radius $R$ due to the deterministic torque.
In particular, the chiral swimming pattern manifests itself in characteristic oscillations at a plateau
for wavenumbers $kR \lesssim 2\pi$ (i.e. $kL\lesssim 4\pi^2M$) and times $t\sim\tau_\omega$, 
which smear out due to rotational diffusion at longer times, $t\gtrsim
\tau_\text{rot}$.
For the case that $\tau_\omega \lesssim\tau_\text{rot}$ (i.e. $M\gtrsim 1$) these oscillations can be rationalized using the approximation of
the pure persistent circular motion
with corresponding ISF, $F(k,t) = J_0(2kR|\sin\omega t /2|)$. In particular,
in this approximation the ISF displays oscillations between unity and the plateau $J_0(2kR)=J_0(kL/\pi M)$.
These oscillations persist for arbitrarily small wavenumbers, yet, the amplitude  $1-J_0(kL/\pi M)\approx(kL/2\pi M)^2$ becomes small.
At infinite P{\'e}clet number, the approximation reproduces the ISF for wavenumbers probing the
radius of the circular motion and for times $t\lesssim\tau_\text{rot}$ (see Fig.~\ref{fig:ISF}~(c) black dashed-dotted line).
The oscillations at a plateau are also predicted by our analytic theory for small wavenumbers $kL\lesssim 5$
and times $t\sim\tau_\omega\lesssim\tau_\text{rot}$ and agree with
the approximate solution. However,
since these oscillations become negligible small, the ISF for small wavenumbers can be approximated by a simple exponential
with effective diffusion coefficient, $\exp(-D_\text{eff}k^2t)$.

For increasing quality factors $M$, the radius of the circular motion decreases ($R=v/\omega=v/(2\pi D_\text{rot} M)$), and
therefore, these characteristic oscillations occur at even larger wavenumbers, which are required to resolve the circular motion.
For times $t\ll\tau_\omega$ the ISF reduces to the ISF of pure persistent swimming motion, $F(k,t) = J_0(vkt)$,
since at these time scales the particle has completed only a small fraction of a circle and, therefore, the motion
appears as a straight line.

Moreover, for increasing quality factor $M$ oscillations at a plateau become stronger at intermediate
times, whereas the effective diffusion of the particle at large length scales
is reduced, and therefore shifts the decaying exponentials
$\exp(-D_\text{eff}k^2t)$ to longer and longer times.  Furthermore, for
decreasing P{\'e}clet numbers these oscillations at a plateau at intermediate times and large
wavenumbers are less pronounced or even smeared out due to the translational
diffusion.

\begin{figure}[htp]
\centering
 \includegraphics[width=\linewidth,keepaspectratio]{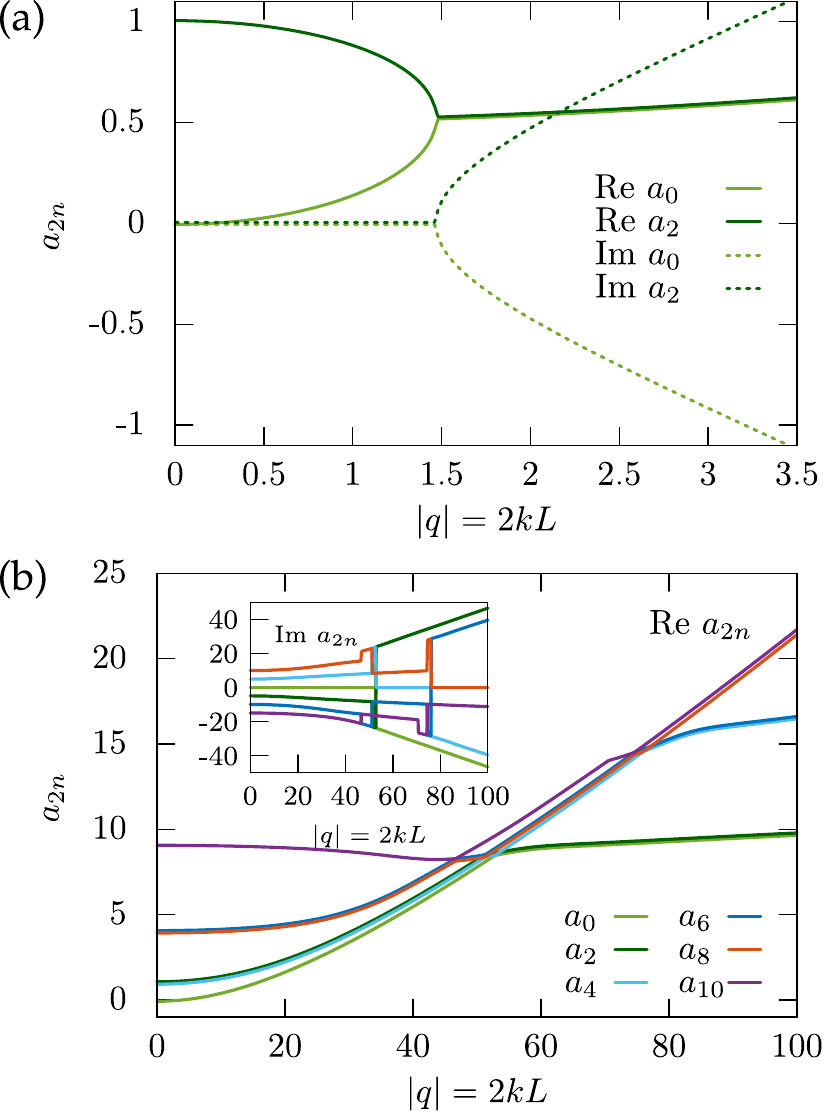}
 \caption{(a) Real (solid line) and imaginary part (dashed line) of the two adjacent eigenvalues $a_0$ and $a_2$
 for  P{\'e}clet number $\text{Pe}=\infty$ and quality factor $M=0$.
(b) Real and imaginary part (inset) of the lowest eigenvalues for  P{\'e}clet number $\text{Pe}=\infty$ and
 quality factor $M=0.8$. The eigenvalues depend continuously on the deformation parameter $|q|=2kL=2kv/D_\text{rot}$, however,
the ordering changes with increasing magnitude.
 \label{fig:eigenvalues}}
\end{figure}
Interestingly, from a mathematical point of view, these oscillations occur
as the operator in Eq.~\eqref{eq:fourier} is not Hermitian and therefore allows for
pairs of complex conjugated eigenvalues. Here, for $M=0$ and infinite
P{\'e}clet number $\text{Pe}=\infty$ we find  that the two lowest neighboring
eigenvalues are real at small wavenumbers, whereas they merge at a certain
point $kL=|q|/2 = 0.75$, where they branch out to a pair of complex conjugates
eigenvalues~\cite{Ziener:2012,Kurzthaler:2016}, see
Fig.~\ref{fig:eigenvalues}~(a). Hence, oscillations in the ISF start to become apparent
for wavenumbers larger than $kL>0.75$, whereas at smaller wavenumbers (e.g.
$kL=0.25$) the eigenvalues are real, and we observe effective diffusion
(compare with Fig.~\ref{fig:ISF}~(a)).

Due to the branching points of the eigenvalues, the ISF depends non-analytically on the wavenumber $kL$,
and the expansion of the ISF in the wavenumber is anticipated to display a finite radius of convergence. Therefore, the oscillations
of the ISF, which become apparent only after the first branching point, cannot be recovered from a perturbation theory
in the wavenumber in terms of the low-order moments.

For $M\neq 0$ we find a more intricate behavior of the lowest
eigenvalues, which contain complex conjugated pairs for all wavenumbers, in contrast to the case of
$M=0$.
In particular, the real parts of these eigenvalues merge with
that of a lower eigenvalue at small wavenumbers, split up at a certain
wavenumber $|q|=2kL$, where they become purely real numbers, and later form a new
pair of complex conjugates with the higher adjacent eigenvalue (see Fig.~\ref{fig:eigenvalues}~(b)).
The eigenvalues depend continuously on $|q|=2kL$, however, they intersect and branch so that
the labeling by increasing magnitude changes.
Nevertheless, the series expansion of the ISF
contains either real eigenvalues or pairs of these complex conjugated
eigenvalues such that the solution always remains real.

\section{Summary and conclusion}
We have elaborated an analytic expression for the ISF of an anisotropic Brownian
circle swimmer in terms of appropriate eigenfunctions and corroborated our
results by stochastic simulations. In addition to oscillations in the
ISF reflecting the persistent swimming motion, the chiral swimming
pattern of the particle manifests itself in oscillations at a plateau at intermediate times and
length scales, where the particle has approximately completed one full circle.
These oscillations, either around zero or at a finite plateau, smear out at large wavenumbers due to the
bare translational diffusion, and at small wavenumbers due to the rotational
diffusion of the particle. In particular, the deterministic torque reduces the
effective diffusion of these circle swimmers at large length scales with respect to a straight
swimmer.

Furthermore, we have computed exact low-order moments of the stochastic process
upon expansion of the ISF in the wavenumber. In particular, we have evaluated
the non-Gaussian parameter, which displays oscillations at intermediate times
mirroring the interplay of circular swimming motion and rotational diffusion of these active agents.
This non-monotonic behavior has also been observed in computer simulations of chiral particles in 3D,
subject to isotropic translational diffusion~\cite{Sevilla:2016}. Similar to the
non-Gaussian parameter of three dimensional anisotropic
particles~\cite{Kurzthaler:2016}, it is positive for short times reflecting the
anisotropic diffusion and approaches zero for long times.

Up to now low-order moments such as the mean-square displacement of these
circle swimmers have been mainly used to extract relevant motility parameters
from experimental observations. 
In these studies, the long- and short-time diffusivities have been compared to experiments
on bimetallic micromotors by Ref.~\cite{Marine:2013}, and the full time dependence
of the mean-square displacement has successfully quantified the dynamics of
bacteria~\cite{Utada:2014} and artificial microswimmers~\cite{Ebbens:2010}.
However, these low-order moments are to a great extend insensitive to the shape of the
probability distribution, whereas more detailed spatiotemporal information on the dynamics of
active particles is encoded in the ISF~\cite{Gardiner:2009}. Only recently, the ISF of a
dilute suspension of the algae \textit{Chlamydomonas reinhardtii} has been
measured in differential dynamic microscopy (DDM) experiments, and displays
characteristic oscillations at a plateau at intermediate times and length scales~\cite{Martinez:2012} as
found within our theory.  To determine the transport properties,
approximations of the ISF for the motion at small length scales have been used.
Here, our analytic theory also predicts the
dynamics of these circle swimmers at larger length scales, where rotational
diffusion starts to play a pivotal role, and the motion of the particle
gets randomized.
%

The analytic expression for the ISF of a Brownian circle
swimmer therefore allows to analyze experimental data of chiral particles
for the full range of length scales. For example, it can be used to extract relevant
motility parameters of anisotropic Janus particles confined between two glass
plates~\cite{Kummel:2013,tenHagen:2014}, or to analyze the role of
diffusion in the circular motion of
bacteria~\cite{Berg:1990,DiLuzio:2005,Lauga:2006,
Hill:2007,Li:2008,DiLeonardo:2011} or sperms~\cite{Woolley:2003,
Riedel:2005,Bohmer:2005,Friedrich:2008} close to surfaces.
Furthermore, the dynamics of a single circle swimmer in a homogeneous environment
presents a suitable starting point to analyze their tactic behavior
as response to an additional external (e.g. gravitational) force~\cite{tenHagen:2014}.
It also serves as a reference to characterize the non-equilibrium behavior of these chiral
particles exposed to spatially heterogeneous media~\cite{Peruani:2013,Schirmacher:2015}.
Similarly, it might be a useful input to establish sorting mechanisms of microswimmers
according to their (chiral) transport properties~\cite{Mijalkov:2013, Chen:2015}.

Due to the mirror symmetry of clockwise and anticlockwise circle swimmers,
the ISF is insensitive to the chirality, and therefore DDM measurements do not allow
detecting the sense of rotation of these active agents.
Hence, to elucidate the interesting question about the role of chirality,
one might measure in the framework of particle tracking
angular correlation functions of different orders.
Furthermore, one can compute analytically and measure
experimentally the low-order moments of the displacements for a fixed initial orientation
in order to test the relative importance of the angular drift with respect to the rotational diffusion.

Moreover, we have also evaluated the ISFs for straight swimmers in two dimensions,
which display qualitatively similar behavior as those in three dimensions~\cite{Kurzthaler:2016},
where the oscillations wash out at small and large wavnumbers due to translational and rotational
diffusion, respectively. Quantitatively, the amplitudes of the oscillations are predicted to be stronger
for particles moving in a plane, which can solely be traced back to the dimension of the system, as
the P{\'e}clet numbers and the relation of the diffusion coefficients are the same.

We anticipate that the analytic expressions for the ISF of anisotropic
particles in three dimensions~\cite{Kurzthaler:2016}, simple run-and-tumble
particles~\cite{Martens:2012}, and Brownian circle swimmers together
permit to discriminate between different swimming behaviors of active
particles, whereas the mean-square displacements are sometimes hardly
distinguishable.  In particular, we have worked out that the occurrence of characteristic
oscillations in the ISF at a plateau constitutes a strong indicator for a chiral swimming
pattern of active particles.

\begin{acknowledgements}
We thank Sebastian Leitmann and Victor Wenin for helpful discussions. This work has been supported by Deutsche Forschungsgemeinschaft (DFG)
via the contract No.~FR1418/5-1 and by the Austrian Science Fund: P~28687-N27.
\end{acknowledgements}

\begin{widetext}
\section*{Appendix}
\appendix

\vspace{-0.1cm}
\section{Stochastic simulation\label{ap:simulation}}
The starting point for the stochastic simulations are the Langevin equations [Eqs.~\eqref{eq:theta}-~\eqref{eq:pos}],
which are discretized according to the Euler scheme~\cite{Gardiner:2009},
\begin{align}
\vartheta(t+\Delta t) 	&= \vartheta(t)+\sqrt{2D_\text{rot}\Delta t} N_\psi +\omega \Delta t\\
\vec{r}(t+\Delta t )	&= \vec{r}(t) + v\vec{u}\Delta t + \left[\sqrt{2 D_\parallel}\vec{u}\vec{u}^T+\sqrt{2D_\perp}(\mathbb{I}-\vec{u}\vec{u}^T)\right]\sqrt{\Delta t} \vec{N}_\xi,
\end{align}
where $\Delta t$ denotes the discretized time step. Here, $N_{\psi}$ and $\vec{N}_\xi$ are
independent and normally distributed random variables with zero mean and unit variance.
To obtain reliable statistics we set the time step $\Delta t = 10^{-3}\tau_\text{rot}$ and
simulate $10^5$ particles.
\section{Equation of motion for the angular correlation function~$C_n(t)$\label{ap:M}}
The equation of motion for the angular correlation function $C_n(t) = \langle\exp\left[\imath n(\vartheta(t)-\vartheta(0))\right]\rangle\equiv\langle c_n(t)\rangle$ can be obtained using It$\bar{\text{o}}$'s Lemma~\cite{Gardiner:2009},
\begin{align}
 \diff  c_n(t) &= \left(\imath n \omega-n^2 D_\text{rot}\right)  c_n(t) \diff t +\imath n  c_n(t) \diff\psi(t),
\end{align}
where $\psi(t)$ denotes a white noise process. Taking the mean, we find immediately the equation of motion for $C_n(t)$
\begin{align}
 \frac{\diff}{\diff t}C_n(t) -nD_\text{rot}(2\pi\imath M-n)C_n(t)&=0.
\end{align}
Alternatively, one can derive this equation by multiplying the Fokker-Planck equation Eq.~\eqref{eq:fourier} for $\vec{k}=0$
with $\exp[\imath n(\vartheta-\vartheta_0)]$, averaging over initial and integrating over final angles, and then integrating by parts.

\section{Non-hermitian eigenvalue problem~\label{appendix:nonhermitian}}
To show that the generalized Mathieu functions are orthogonal in the sense of Eq.~\eqref{eq:norm},
we first define the scalar product for $\pi$-periodic functions $\varphi(x), \psi(x)$ by
\begin{equation}
 \langle \varphi |  \psi \rangle = \frac{1}{\pi}\int_{0}^{\pi} \varphi(x)^* \psi(x) \diff x.
\end{equation}
The adjoint operator of the Sturm-Liouville operator in Eq.~\eqref{eq:Sturm} with respect to this scalar product fulfills
\begin{align}
L^+ = L(q,c,M)^+ = -\frac{\diff^2}{\diff x^2}+2q^*\cos(2x)+c^2 \cos^2(2x)- 4\pi M\frac{\diff}{\diff x}.
\end{align}
Here, we recall that generally if $r_m(q,c,M,x)$ is a right-eigenfunction with eigenvalue $a_m(q,c,M)$, $Lr_m = a_m r_m$, and $l_n(q,c,M,x)$ a left-eigenfunction
with eigenvalue $b_n(q,c,M)$,
$L^+l_n = b_n^* l_n$, then one finds
\begin{align}
& \langle l_m | L r_n \rangle = a_n \langle l_m | r_n \rangle  =\langle L^+ l_m | r_n \rangle = b_m \langle l_m | r_n \rangle,
\end{align}
and therefore
\begin{equation}
 (a_n - b_m) \langle l_m | r_n \rangle = 0.
\end{equation}
Then eigenfunctions corresponding to different eigenvalues are mutually orthogonal.
If the eigenvalue is the same, we label the eigenfunctions
to eigenvalue $a_m$ by $\langle l_m|$ and $|r_m\rangle$.

A direct calculation shows that
\begin{align}
[ L^+(q,c,M) l_n(q,c,M,x) ]^*  &= a_n(q,c,M) l_n(q,c,M,x)^* \nonumber \\
&= \left[-\frac{\diff^2}{\diff x^2} + 2q \cos(2x) + c^2 \cos^2(2x) -4 \pi M \frac{\diff}{\diff x} \right] l_n(q,c,M,x)^*.
\end{align}

A change of variables $x \mapsto -x$ then  yields
\begin{align}
L(q,c,M)l_n(q,c,M,-x)^* = a_n(q,c,M) l_n(q,c,M,-x)^*, \label{eq:L}
\end{align}
and we conclude that up to normalization $l_n(q,c,M,x)^* \propto r_n(q,c,M,-x)$.
Hence, we choose $l_n(q,c,M,x)^* = r_n(q,c,M,-x)$ and normalize the set of eigenfunctions by $\langle l_n | r_m \rangle = \delta_{nm}$,
which corresponds to the orthogonality relation in Eq.~\eqref{eq:norm}.

Furthermore, due to the equality of the eigenvalues $b_n=a_n$ we find
\begin{align}
 L^+(q^*,c,-M)l_n(q^*,c,-M,x)
&=a_n(q^*,c,-M)^*l_n(q^*,c,-M,x),
\end{align}
and as $L(q,c,M)=L^+(q^*,c,-M)$, we observe by comparison with Eq.~\eqref{eq:L} that the
eigenvalues fulfill the symmetry relation $a_n(q,c,M) = a_n(q^*,c,-M)^*$.

In addition, for purely imaginary $q$, similar to the discussion in the main text, $z(q, c, M, x+\pi/2)$ is eigenfunction to $L(-q, c, M)$ with eigenvalue $a(q, c, M)$,
and therefore the spectrum does not depend on the sign of the velocity.

\section{Mean-quartic displacement~\label{appendix:meanquartic}}
To compute the mean-quartic displacement, we expand the ISF up to the fourth order in the wavenumber
$\mathcal{O}(k^4)$. Therefore, we consider a Dyson series [Eq.~\eqref{eq:dyson}] with perturbations including the order
$\mathcal{O}(\hat{V}_\text{swim}^4,\hat{V}_\text{diff}^2,\hat{V}_\text{swim}^2\hat{V}_\text{diff},\hat{V}_\text{diff}\hat{V}_\text{swim}^2,\hat{V}_\text{swim}\hat{V}_\text{diff}\hat{V}_\text{swim})$.
Using the properties of the matrix elements $\langle m|\hat{V}|n\rangle$ [Eq.~\eqref{eq:vmn}], which are non-zero only for $m=n, n\pm1,n\pm2$,
the expansion of the ISF can be determined by solving for a finite number of multifold time-integrals.
Since we only need the zeroth matrix element of the time-evolution operator [Eq.\eqref{eq:ISFexpanded}],
it suffices to consider the modes $m=0,\pm1,\pm2$ only, and we obtain the solution by a linear combination of the five modes: $1$, $\exp(-D_\text{rot}t) \sin(\omega t)$, $\exp(-D_\text{rot}t) \cos(\omega t)$,
$\exp(-4D_\text{rot}t)\sin(2\omega t)$, and $\exp(-4D_\text{rot}t)\cos(2\omega t)$.
The computations are lengthy and have therefore been implemented and evaluated in a computer algebra system~\cite{Mathematica}.

Comparing the expansion of the ISF to Eq.~\eqref{eq:sincex}, we obtain the mean-quartic displacement of the Brownian circle swimmer,

\begin{align}
  & \langle|\Delta\vec{r}(t)|^4\rangle = \frac{e^{-4 D_\text{rot} t}}{\omega  \left(D_\text{rot}^2+\omega ^2\right)^4 \left(36 D_\text{rot}^4+13 \omega ^2 D_\text{rot}^2+\omega ^4\right)^2} \Bigl\{2 \omega  \Bigl[\left(v^2+2 D_\text{rot} \Delta D\right) \omega ^2-\Delta D \omega ^3-D_\text{rot} \left(9 D_\text{rot} \Delta D-5 v^2\right) \omega\notag\\
  &-6 D_\text{rot}^2 \left(v^2-3 D_\text{rot} \Delta D\right)\Bigr] \Bigl[18 \Delta D D_\text{rot}^3+\left(9 \Delta D \omega -6 v^2\right) D_\text{rot}^2+\omega  \left(2 \Delta D \omega -5 v^2\right) D_\text{rot}+\omega ^2 \left(v^2+\Delta D \omega \right)\Bigr] \left(D_\text{rot}^2+\omega ^2\right)^4\cos (2\omega t  ) \notag\\
  &-4 \omega ^2 \left(9 \Delta D D_\text{rot}^2-5 v^2 D_\text{rot}+\Delta D \omega ^2\right) \Bigl[6 \left(3 D_\text{rot} \Delta D-v^2\right) D_\text{rot}^2+\left(v^2+2 D_\text{rot} \Delta D\right) \omega ^2\Bigr] \left(D_\text{rot}^2+\omega ^2\right)^4\sin (2\omega  t ) \notag\\
  &+2 e^{4 D_\text{rot} t} \omega  \left(9 D_\text{rot}^2+\omega ^2\right)^2 \Bigl[\left(4 D_\text{rot} t+1\right) \Delta D^2 \omega ^{10}+\Bigl(\left(2 D_\text{rot} t+1\right) \left(2 D_\text{rot} t+3\right) v^4+6 D_\text{rot} \left(4 D_\text{rot} t+3\right) \Delta D v^2\notag\\
                                     &+32 D_\text{rot}^3 t \Delta D^2\Bigr) \omega ^8+D_\text{rot}^2 \Bigl(\left(4 D_\text{rot} t (10 D_\text{rot} t+21)+27\right) v^4+6 D_\text{rot} \left(28 D_\text{rot} t+5\right) D_\text{rot} \Delta D v^2+2 D_\text{rot}^2 \left(44 D_\text{rot} t-5\right) \Delta D^2\Bigr) \omega ^6\notag\\
                                     &+D_\text{rot}^4 \Bigl(3 \left(4 D_\text{rot} t (11 D_\text{rot} t+16)-61\right) v^4+18 D_\text{rot} \left(20 D_\text{rot} t-9\right) \Delta D v^2+4 D_\text{rot}^2 \left(28 D_\text{rot} t-5\right) \Delta D^2\Bigr) \omega ^4\notag\\
                                     &+D_\text{rot}^6 \Bigl(\left(4 D_\text{rot} t (40 D_\text{rot} t-31)-723\right) v^4+6 D_\text{rot} \left(52 D_\text{rot} t-57\right) \Delta D v^2+D_\text{rot}^2 \left(68 D_\text{rot} t-15\right) \Delta D^2\Bigr) \omega ^2\notag\\
                                     &+16 \bar{D}^2 t^2 \left(D_\text{rot}^2+\omega ^2\right)^4 \left(4 D_\text{rot}^2+\omega ^2\right)^2+4 D_\text{rot}^8 \Bigl(\left(4 D_\text{rot} t (4 D_\text{rot} t-15)+87\right) v^4+6 D_\text{rot} \left(4 D_\text{rot} t-7\right) \Delta D v^2\notag\\
                                     &+D_\text{rot}^2 \left(4 D_\text{rot} t-1\right) \Delta D^2\Bigr)+16 \bar{D} t v^2 \left(D_\text{rot}^2+\omega ^2\right)^2 \left(4 D_\text{rot}^2+\omega ^2\right)^2 \left((D_\text{rot} t-1) D_\text{rot}^2+(D_\text{rot} t+1) \omega ^2\right)\Bigr]\notag\\
                                     &+8 e^{3 D_\text{rot} t} v^2 \left(4 D_\text{rot}^2+\omega ^2\right)^2 \Bigl[-\omega  \Bigl(9 \left((6 D_\text{rot} t+49) v^2-24 D_\text{rot} \Delta D\right) D_\text{rot}^8-4 \left((39 D_\text{rot} t+259) v^2+96 D_\text{rot} \Delta D\right) \omega ^2 D_\text{rot}^6\notag\\
                                     &-2 \left((108 D_\text{rot} t-31) v^2+56 D_\text{rot} \Delta D\right) \omega ^4 D_\text{rot}^4+4 \left((1-D_\text{rot} t) v^2+16 D_\text{rot} \Delta D\right) \omega ^6 D_\text{rot}^2\notag\\
                                     &+\left((2 D_\text{rot} t+1) v^2+8 D_\text{rot} \Delta D\right) \omega ^8-4 \bar{D} t (D_\text{rot}-\omega ) (D_\text{rot}+\omega ) \left(D_\text{rot}^2+\omega ^2\right)^2 \left(9 D_\text{rot}^2+\omega ^2\right)^2\Bigr) \cos (\omega t  )\notag\\
                                     &-\Bigl((8 \bar{D} D_\text{rot} t-\Delta D) \omega ^{10}+D_\text{rot} \left(160 \bar{D} t D_\text{rot}^2+11 \Delta D D_\text{rot}+(12 D_\text{rot} t+5) v^2\right) \omega ^8+2 D_\text{rot}^3 \bigl(472 \bar{D} t D_\text{rot}^2+107 \Delta D D_\text{rot}\notag\\
                                     &+10 (5 D_\text{rot} t+2) v^2\bigr) \omega ^6 +2 D_\text{rot}^5 \left(720 \bar{D} t D_\text{rot}^2+155 \Delta D D_\text{rot}+(265-46 D_\text{rot} t) v^2\right) \omega^4+3 D_\text{rot}^7 \bigl(216 \bar{D} t D_\text{rot}^2+9 \Delta D D_\text{rot}\notag\\
                                     &-20 (3 D_\text{rot} t+16) v^2\bigr) \omega ^2+81 D_\text{rot}^9 (v^2-D_\text{rot} \Delta D)\Bigr) \sin (\omega t  )\Bigr]\Bigr\}. \label{eq:mqd}
\end{align}

\section{Numerical evaluation of the eigenfunctions~\label{numerics}}
To numerically evaluate the ISF [Eq.~\eqref{eq:ISFanalytics}], the eigenvalues $a_{2n}$ and the integrals
of the eigenfunctions $\text{ee}_{2n}(q,c,M,\vartheta/2)$ are needed.
Therefore, we insert the expanded eigenfunction [Eq.~\eqref{eq:fm}] into the eigenvalue problem [Eq.~\eqref{eq:Sturm}] and obtain
the relation for the Fourier coefficients
\begin{align}
   \left(a_{2n}-\alpha_{2m}\right) A_{2m}^{2n} -q \left(A_{2m+2}^{2n}+A_{2m-2}^{2n}\right)-\frac{c^2}{4} \left(A_{2m+4}^{2n}+A_{2m-4}^{2n}\right)&= 0,
\end{align}
with coefficients $\alpha_m =8\pi\imath M m+4m^2+c^2/2$. For $c=M=0$ the relation for the Fourier coefficients is identical to
that of the conventional even and odd Mathieu functions except for the zeroth mode~\cite{NIST:print,NIST:online}.
To compute numerically the Fourier coefficients and eigenvalues, we solve the eigenvalue problem
$\textsf{M}\vec{A}^{2n} =a_{2n}\vec{A}^{2n}$, where the eigenvector contains the Fourier coefficients of the expansion,
$\vec{A}^{2n}=\left[\ldots,A^{2n}_{-2},A^{2n}_0,A^{2n}_2,\ldots\right]^T$, and the matrix~$\sf{M}$ is a band matrix with diagonal elements $M_{mm}=\alpha_{2m}$,
and off-diagonals $M_{m,m-1}=M_{m,m+1}=q$, and $M_{m,m-2}=M_{m,m+2}=c^2/4$.

The orthonormalization of the eigenfunction translates in the matrix representation to  $\sum_{j=-\infty}^\infty A_{2j}^{2m}A_{2j}^{2n} = \delta_{mn}$.
To finally evaluate the ISF, we compute the integrals in Eq.~\eqref{eq:ISFanalytics},
which reduce to the zeroth Fourier coefficient,  
$\int_0^{2\pi}\diff\vartheta \ \text{ee}_{2n}(q,c,M,\vartheta/2)= 2\pi A_0^{2n}$.
In practice, the matrix is truncated at an appropriate dimension such that the normalization of the ISF at $t=0$
is fulfilled, $\sum_{n=-\infty}^\infty A_0^{2n}A_0^{2n}=1$, with reasonable
accuracy.

\end{widetext}



\bibliography{swimmer} 

\end{document}